Site-dependent evolution of electrical conductance from tunneling to atomic point contact

Howon Kim and Yukio Hasegawa*

The Institute for Solid State Physics, The University of Tokyo

5-1-5, Kashiwa-no-ha, Kashiwa 277-8581, Japan

Abstract

Using scanning tunneling microscopy (STM), we investigated the evolution of electrical conductance between a Pb tip and Pb(111) surface from tunneling to atomic point contact at a site that was defined with atomic precision. We found that the conductance evolution depended on the contact site, for instance, on-top, bridge, or hollow (hcp and fcc) sites in the Pb lattice. In the transition from tunneling to contact regimes, the conductance measured at the on-top site was enhanced. In the point contact regime, the hollow sites had conductances larger than those of the other sites, and between the hollow sites, the hcp site had a conductance larger than that of the fcc site. We also observed the enhancement and reversal of the apparent height in atomically resolved high-current STM images, consistent with the results of the conductance traces. Our results indicate the importance of atomic configuration in the conductance of atomic junctions and suggest that attractive chemical interactions have a significant role in electron transport between contacting atoms.

The properties of electron transport through atomic-size contacts are of fundamental interest in view of the potential nanoscale device technologies [1]. The challenges of atomic point contacts have been addressed by various approaches, including the break-junction method [2-5], transmission electron microscopy [6], and scanning tunneling microscopy (STM) [7-15], as well as in theoretical calculations [16-22]. Several seminal phenomena, such as quantization in conductance [3,8], conductance through a single molecule [4,11,13], atom transfer [9], and atomically controlled Kondo interactions [10,14], have been observed and investigated. In these experiments, the evolution of conductance was monitored during the repeated contact formation between the two electrodes, and the captured traces of the conductance as a function of the gap separation were analyzed statistically by using conductance histograms [5,13]. The histograms taken on metals, except noble ones, exhibit broad distributions with multiple peaks, which have been attributed to the stochastic nature of the contact formation. In fact, theoretical studies have indicated that the lateral configuration of the contact-forming atoms causes this significant variation in conductance [16-22]. However, direct experimental evidence proving this variation has not been obtained yet; despite the unprecedented atomic-resolution imaging possible with STM, the ultimate lateral resolution has not been utilized for point contact formation. In this Letter, we report on atomic contacts made with lateral atomic precision. Contacts were formed at on-top, bridge, and hollow sites in the crystallographic lattice of the substrate surface, as shown schematically in Fig. 1(a), by defining the contact sites in atomically resolved STM images taken prior to contact formation. We investigated the site dependence of electrical conductance from tunneling to contact, and found that the conductance evolution clearly depended on the contact site. Broad and multiple peaks were reproduced in the conductance histograms, and each peak could be assigned to a specific atomic contact configuration, which indicates the importance of precise control of atomic geometry in determining point contact conductance.

The experiments were performed using a cryogenic ultrahigh vacuum STM (Unisoku, USM-1300S) with a controller (Specs, Nanonis) [23]. By advancing a Pb-covered tip toward a flat-top fcc Pb island structure formed on a Si(111) substrate (As-doped, 1-3mΩ·cm), we formed atomic-size junctions at a temperature of 2.1K. To fabricate the Pb-covered tip, a PtIr tip was intentionally indented into a Pb island. The Pb-Pb junction was confirmed from tunneling spectra that exhibit gaps whose widths corresponded to the sum of the two superconducting electrodes. The electrical conductance was obtained by dividing the measured tip current by the bias voltage $V_S$ (3.8mV) that was applied on the substrate and was much larger than the gap width (~2.5mV).

A trace of the electrical conductance as a function of the tip displacement $\Delta z$ was measured by advancing/retreating the tip toward/from the surface after turning off the STM feedback. In the experiments shown in Figs. 1 and 2, $\Delta z$ was measured from the tip height set with the tunneling current $I_t$ of 30nA at $V_S$=3.8 mV. Conductance traces were acquired at every point during the scanning of the tip over the surface. From an STM image acquired simultaneously (Fig. 2(a)), the conductance traces measured at specific sites in the atomic lattice of the substrate, specifically, at on-top, bridge, and hcp/fcc hollow sites, were obtained. Pb islands were formed on a Pb-induced $\sqrt{3}\times\sqrt{3}$-reconstructed structure, and their azimuthal directions were rotated with respect to the Si substrate to form moiré patterns with various rotational angles and periodicities [13,24,25]. The two hollow sites, namely the hcp and fcc sites, which respectively has and does not have an atom in the second layer, could be identified from the lateral relation between the atomic lattices of the two surface layers separated by a monolayer-high step, whose details are provided in Section I of the supplemental materials (SM) [26].

Figure 1(b) shows the conductance traces taken at on-top, bridge, fcc, and hcp sites on the surface of a seven-monolayer (ML) Pb island. For each plot, 10 traces were obtained from the corresponding sites marked in the inset STM image and averaged. All 100 traces in the inset STM image are plotted in the gray line in the zoomed plot (right panel). For 0pm<$\Delta z$<100pm, the conductances measured at all the sites follow the same exponential line (straight in the semi-log plot), indicative of the tunneling nature of the conductance. At closer distances (-40pm<$\Delta z$<0pm) the conductance values increase more steeply than the exponential line. Then, by further advancing the tip toward the surface ($\Delta z$<-45pm), the conductance traces are gradually suppressed around the quantum conductance $G_0$, which implies the point contact regime. $G_0$ is given by $2e^2/h$ (~77.5μS), where -$e$ is the electron charge and $h$ is Planck's constant. These features of the traces are consistent with those reported in point contact studies performed on adsorbed atoms [10,12], molecules [11], and flat metal surfaces without site specification [12,13].

While the overall shapes are similar, the conductances measured at different sites clearly evolve in different manners. Around Δz=-27pm, which is indicated by "Br" in the right panel of Fig. 1(b), the conductance increases more steeply for the on-top site than it does for the other sites, with its trace branching off. Because of the branching, the sequence of the conductances at Δz<-27 pm is on-top>bridge≈hollow. Around -40~-45pm, the conductance traces cross over, which is indicated by "Cr". At distances closer than the crossover point, that is, in the contact regime, the conductance sequence is reversed as hollow>bridge>on-top. These results represent the first observations of the branching and enhancement of the conductance at an on-top site and of the crossover and reversal between the on-top and hollow sites in the evolution of conductance, which were made possible by the site-specific conductance measurements. Note that in these measurements the traces did not show hysteresis; a trace taken during the tip retreat is basically the same as that taken during approach (see SM Sec.II). We carried out conductance trace measurements with several tips and Pb islands and observed similar behaviors (another example is shown in Fig. 3).

In order to spatially demonstrate the site dependence and conductance sequences mentioned above, we performed real-space mappings of the conductance in the on-top conductance-enhancement region (between "Br" and "Cr") and in the contact regime. Figure 2(a) is an STM image showing the atomic contrast taken simultaneously with 64×64 conductance traces, including those shown in Fig. 1(b). At a tip displacement $\Delta z$ of -32pm, which is marked with a vertical line in Fig. 1(b), the conductance mapping (Fig. 2(b)) exhibits bright contrast at the on-top site, similarly to that in the topographic image. As the conductance mapping at $\Delta z=0$ does not have any contrast, the bright contrast indicates the conductance enhancement at the on-top site. On the other hand, the conductance mapping in the point contact regime (Fig. 2(c), $\Delta z=-50$pm) has its contrast reversed from that of the topographic image, indicating a larger conductance at the hollow site than at the on-top site.

In the topographic image, a slight corrugation (3-4pm) due to the moiré structure, whose period is ~3.7nm [13, 24, 25], was observed (see SM Sec.III for more information). The moiré induces small but significant variations in the conductance values, as seen in the conductance mappings of Figs. 2(b) and 2(c). To avoid these influences, we obtained the conductance traces from the high-moiré area (boxed in Fig. 2(a)) for the analysis shown in Fig. 1.

From the site-dependent conductance traces, a histogram of the conductance can be obtained. Figure 2(d) shows the conductance distribution at $\Delta z$=-50pm. Several peaks were observed in the histogram, and each peak was assigned to the conductance of a specific site, as shown in the figure. The distribution of the histogram indicates that variation in the contact site is one of the causes of the conductance variations observed in the atomic point contact measurements. As one may already notice in Figs. 1 and 2, we also found differences in the point contact conductances between the two hollow sites, the hcp and fcc sites. The conductance traces (Fig. 1(b)) and conductance image at $\Delta z$=-50pm (Fig. 2(c)) clearly show a larger conductance at the hcp site than at the fcc site. Thus, the atomic point contact conductance is quite sensitive to the atomic configuration of the contact.

Since the conductance at the hollow site was considerably larger than that at the on-top site in the point contact regime, one could expect an apparent height reversal between the two sites in the topographic STM images, if taken in the regime. Figure 3 shows STM images (a) and their cross-sectional plots (b) taken with various currents within a single frame of scanning ($V_S$=9.3mV). The current was set quite high, from 100nA to 600nA, compared with that used in standard STM imaging. With set currents of 100nA ($0.14G_0$) to 400nA ($0.55G_0$), typical atomically resolved images were obtained. In the images corresponding to 200nA ($0.28G_0$) and 300nA ($0.42G_0$), the atomic corrugation was enhanced, as the conductance corresponds to the region in which the on-top site conductance is larger than that of the hollow site, as shown in the conductance traces of Fig. 3(c). In the images obtained with 500nA ($0.69G_0$) and 600nA ($0.83G_0$), which correspond to distances closer than the crossover point of Fig. 3(c), the atomic contrast was reversed. We confirmed that the observed contrast is reversible, as shown in Sec.VI of SM.

The observed contrast variations can be explained by the conductance traces shown in Fig. 3(c), which were taken prior to the STM imaging. The horizontal lines correspond to the conductances at which the STM images were taken. In this plot, the conductance traces of the hollow site were shifted by -2pm in order to compensate for the height difference at $\Delta z=0$ ($I_t=100$nA), where the conductance trace measurements were started (see SM Sec.IV for details). The height difference at $\Delta z=0$ was obtained from the 0.14$G_0$ STM image shown in Fig. 3(a). By accounting for the height difference at $\Delta z=0$, Fig. 3(c) directly explains the apparent contrast difference $\Delta d$ between the on-top and hollow sites in the STM images taken with a given conductance. In Fig. 3(d), $\Delta d$ in the STM image (Fig. 3(a)) is compared with that estimated from the conductance traces (Fig. 3(c)), demonstrating the consistency of the two measurements. Similar reversed STM images were also reported on Pb/Ag(111) with a Pb tip [13], although corresponding site-specific conductance traces were not presented.

First-principles calculations of the site-specific conductances from the tunneling to atomic point contact on close-packed metal surfaces have been performed by several groups [17,18,20-22], originally in order to investigate why STM images taken on the surfaces have larger corrugations than those expected from the local density of states distributions. In the calculations, the conductance enhancement at the on-top site, conductance crossover, and reversal between the on-top and hollow sites were predicted, as observed in our experiments, regardless of the metal elements involved. In the calculations, the attractive chemical interaction between the surface and tip apex atoms and the resulting relaxation in their atomic positions play significant roles in the development of the conductance, as schematically depicted in Fig. 4. When the tip approaches the surface from the tunneling regime (i), the attractive force between the tip apex atom and a surface atom operates first at an on-top site (red atoms in (ii)) because the distance between atoms is smaller than it is at hollow sites, making the conductance deviate from those at the other sites. Then at closer distances, crossover occurs in the attractive force as it does in the conductance. In the contact regime, the attractive force becomes stronger at the hollow sites because the number of involved atoms (red atoms in (iii)) is greater than that at the on-top site, and, therefore, the conductance becomes larger. Several possible causes of the increased conductance related to the attractive interaction have been discussed, such as modification of the electronic states, collapse of the tunneling barrier, and formation of ballistic channels due to bond formation, but its mechanism has not been clarified yet as supportive experimental results have been lacking. It has been pointed out that in the case of Pb-Pb point contacts, the spin-orbit coupling may also play a significant role in the conductance [16]. Theoretical studies including this effect are, thus, needed to understand the mechanism.

Concerning the chemical interaction between the tip apex atom and surface atoms, non-contact atomic force microscopy (nc-AFM) has been utilized to obtain the site-dependent force evolution on metal surfaces, but results showing distinctive site dependence have not been reported yet because significant background due to the van der Waals force obscures the chemical dependence. Recently, contrast-reversed topographic images were reported for a Cu(111) surface using nc-AFM at small distances [22], proving the force reversal predicted by the first-principles theories. Simultaneous current and force measurements have also been performed using nc-AFM, but so far atomically resolved site-specific measurements have been successful only on semiconductor surfaces [29], whose atomic spacing is larger than that of metal surfaces. Combined with force-sensitive AFM results, precise site-specific conductance traces will provide a key to understanding the relation between electron transport and chemical interaction in the fundamental structures.

As mentioned above, we observed the larger contact conductance at an hcp site than at an fcc site, which was the first experimental observation, while it was predicted by the theory and presumably the chemical interaction again contributes to this difference [18]. In fact, the fcc-hcp contrast has been observed in atom manipulation images [30,31], in which an STM image was taken with a single adsorbed atom dragged or manipulated by a tip during the scanning. We also observed the fcc-hcp contrast with a single Pb atom manipulated with a tip, as shown in Sec.V of SM. Since the experimental conditions and image appearances of the two methods are quite different, however, we can safely rule out the possibility that the observed fcc-hcp contrast shown in Fig. 2(c) is due to the atom manipulation images.

In summary, using low-temperature STM, we measured site-specific conductance traces from tunneling to point contact between a Pb tip and Pb surface and for the first time observed an enhanced conductance at an on-top site in the transition and conductance reversal in the point contact regime. The obtained high-current STM images also show them in the contrast of the atomically resolved images. Our results clearly indicate the importance of the atomic configuration in the conductance of point contacts. Chemical interactions between the contact-forming atoms presumably play a significant role in the electron transport of the ultimately squeezed systems.

The authors are grateful to Profs. Akira Sakai and Ruben Perez for fruitful discussion. This work is partially funded by Grants-in-Aid for Scientific Research, Japan Society for the Promotion of Science (21360018, 25286055).

References

*corresponding author: hasegawa@issp.u-tokyo.ac.jp

Figure captions

Fig. 1: (color online) (a) Schematic of the conductance measurements as a function of the tip displacement $\Delta z$. (b) Conductance traces measured at on-top (black), bridge (green), hcp (red), and fcc (blue) sites on a flat Pb(111) surface, which were obtained from each 10 points marked in the inset STM image (1.2×1.2nm$^2$, $V_S$=3.8mV, $I_t$=30nA) and averaged. The right panel is a zoom of the dotted area in the left panel. All 100 traces (gray) taken in the inset image are also plotted. The circles labeled "Cr" and "Br" indicate the regions in which the crossover and branching of the conductance traces occur, respectively. The two vertical lines correspond to the distances at which the conductance mappings shown in Figs. 2(b) and 2(c) were taken.

Fig. 2: (color online) Spatial mappings of the conductance at various tip displacements (a) topographic image ($V_S$=3.8mV, $I_t$=30nA, 3.0×3.0nm$^2$) taken simultaneously with 64×64 conductance traces. Dotted boxes indicate the area analyzed in Fig. 1. (b) $\Delta z$=-32pm (between "Br" and "Cr" in Fig. 1(b)) (c) $\Delta z$=-50 pm (contact regime) (d) histogram of the conductance at $\Delta z$=-50pm. The arrows correspond to the averaged conductance values taken at the four sites. The solid red line is a visual guide.

Fig. 3: (color online) (a) High-current STM images ($V_S$=9.3mV, 5.0×5.0nm$^2$). The set current was changed from 100nA (bottom) to 600nA (top) in increments of 100nA. The offset height of each section was adjusted so that the atomic contrast could be clearly seen. (b) Cross-sectional plots along the lines drawn in each section of the STM image (a). Long-range corrugation due to the moiré was eliminated by FFT filtering. Each plot was offset for clarity. (c) Conductance traces measured at the on-top and hollow sites prior to the STM imaging shown in (a). The trace of the hollow site was shifted by -2pm in order to compensate for the height difference at $\Delta z$=0 ($V_S$=9.3mV, $I_t$=100nA). The horizontal lines correspond to the conductances at which the STM images were taken. (d) The height difference between the on-top and hcp sites obtained from the STM images and conductance traces. Positive (negative) values mean higher (lower) contrast at the on-top site than at the hollow site.

Fig. 4: Schematics showing the configuration of the tip and surface atoms from tunneling to point contact at on-top (upper panel) and hollow (lower panel) sites. The atoms whose positions are relaxed by the chemical interaction are colored red. Note that the amount of the tip displacement is exaggerated and not in scale.

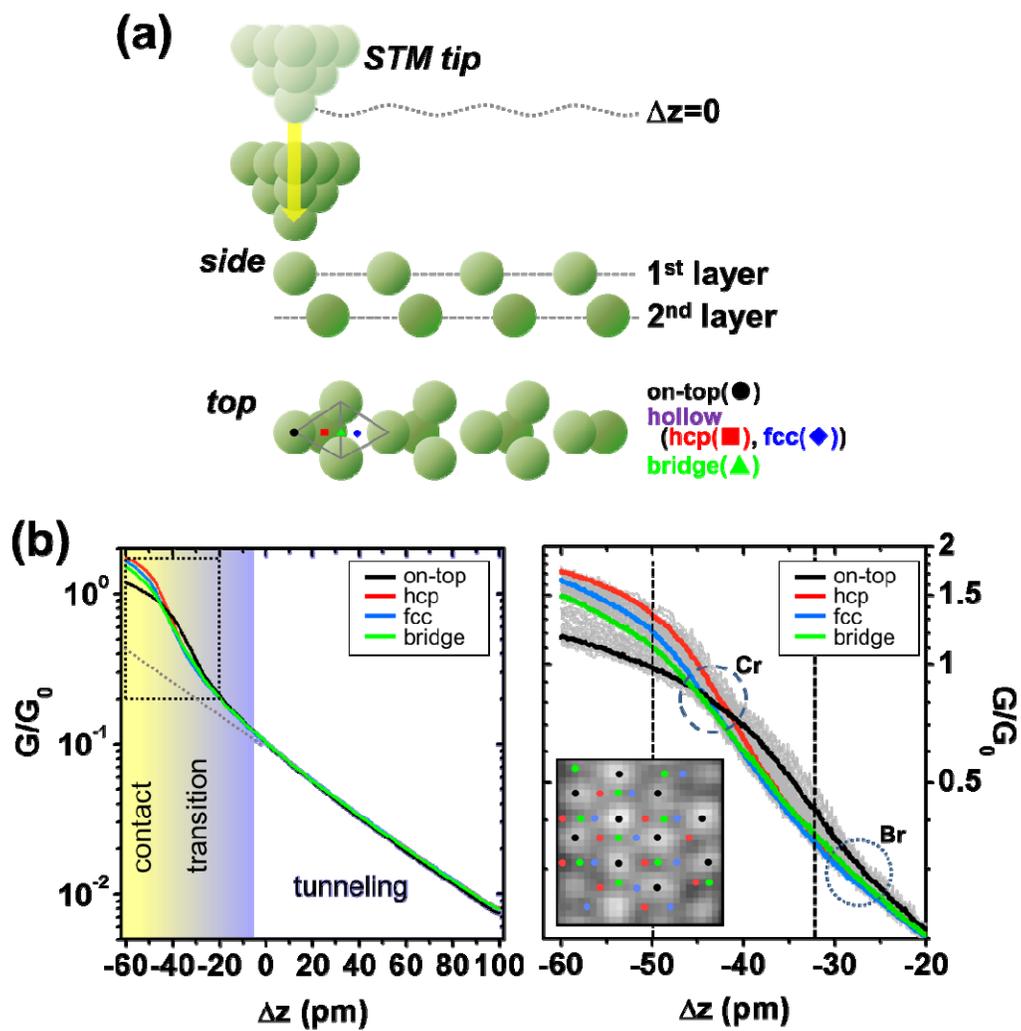

Fig. 1

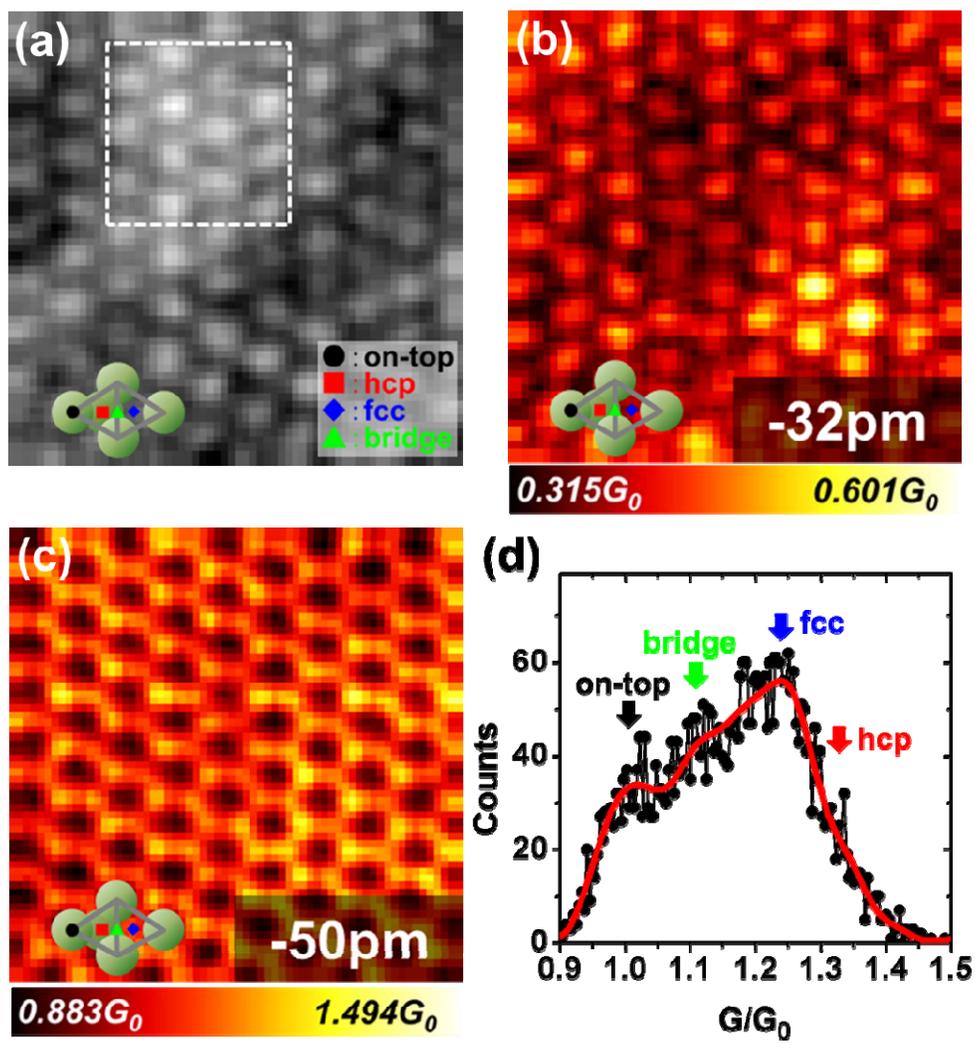

Fig. 2

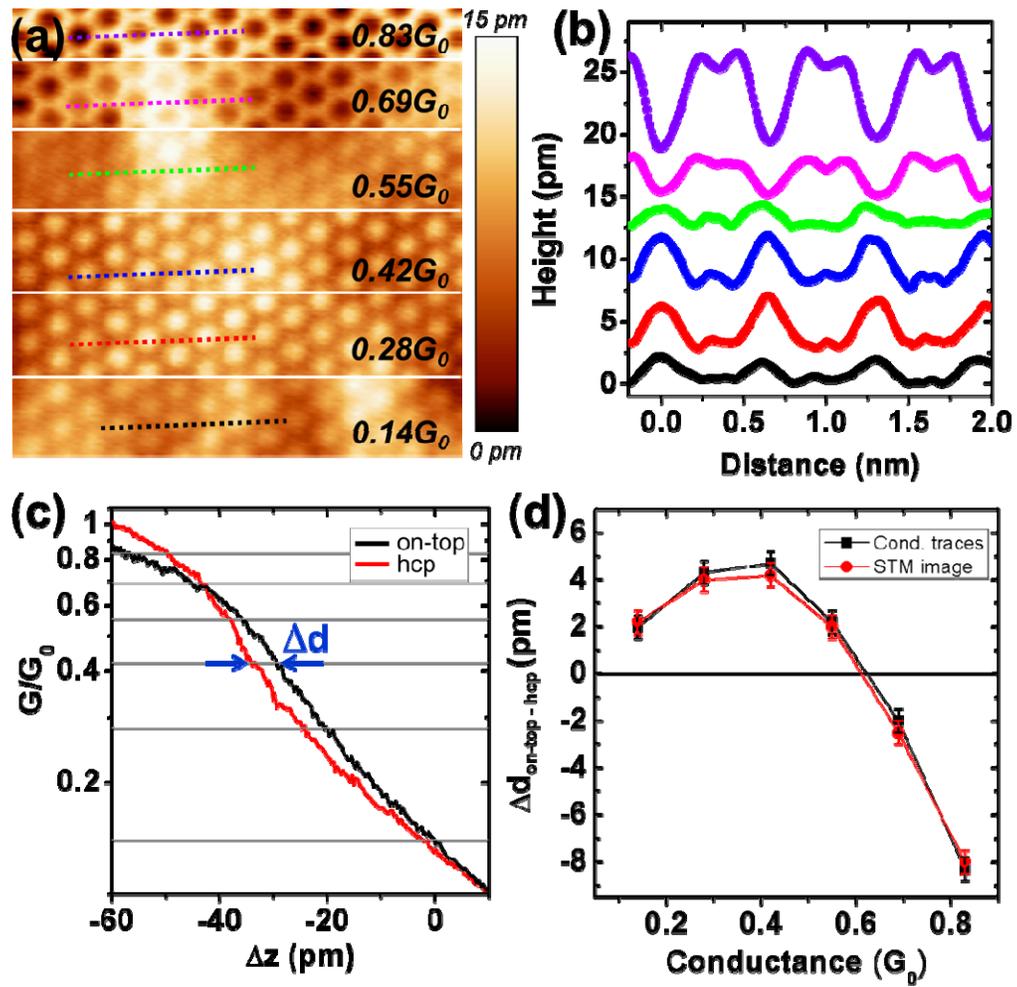

Fig. 3

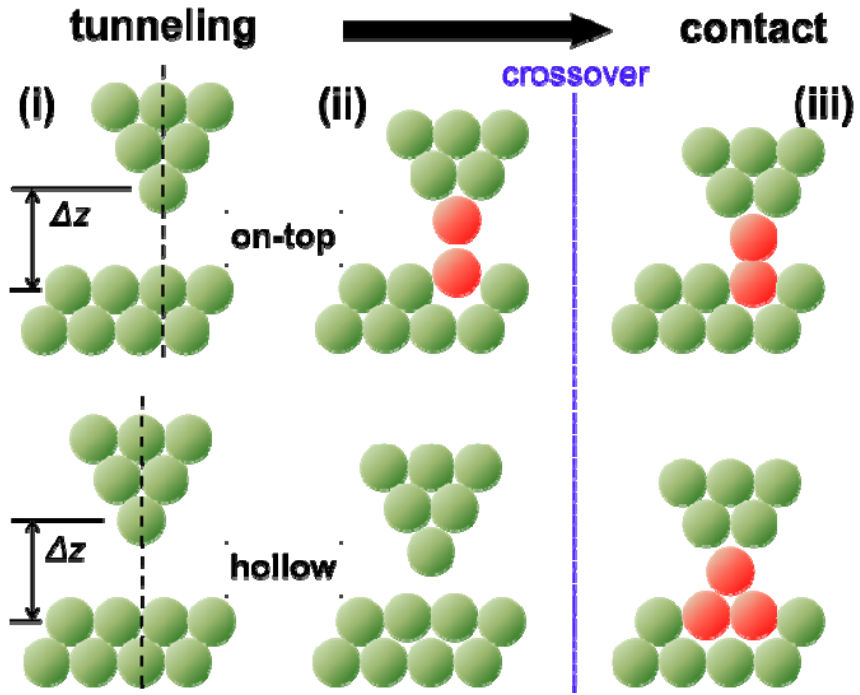

Fig. 4

**Supplemental materials**

of "Site-dependent evolution of electrical conductance from tunneling to atomic point contact"

by Howon Kim and Yukio Hasegawa

### I. Assignment of the fcc and hcp sites in the Pb(111) surface

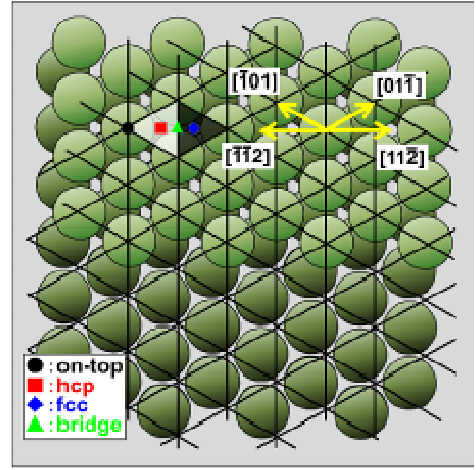

Fig. S1: atomic structural model of the fcc surface with a monoatomic step

The fcc and hcp sites in the (111) surface of fcc structure can be identified from lateral shift of the atomic lattices between the topmost surface layer and the layer beneath the topmost one. Figure S1 shows the relation of the atomic lattices of two surface layers and the fcc/hcp sites in the lattice. In order to investigate the lattice shift, we obtained a single-frame STM image showing atomic resolution on two terraces separated by a monoatomic-high step structure, as shown in Fig. S2. Since the atomic lattice of the lower layer (bottom in Fig. S2) was shifted to the left with reference to the upper terrace lattice, the fcc/hcp sites were determined as depicted in Fig. S2(a).

Since atom manipulation images taken in this area were found to show bright contrast in the hcp sites (Fig. S2(b)), we used the atom manipulation images as a reference to determine the fcc and hcp sites on other Pb islands.

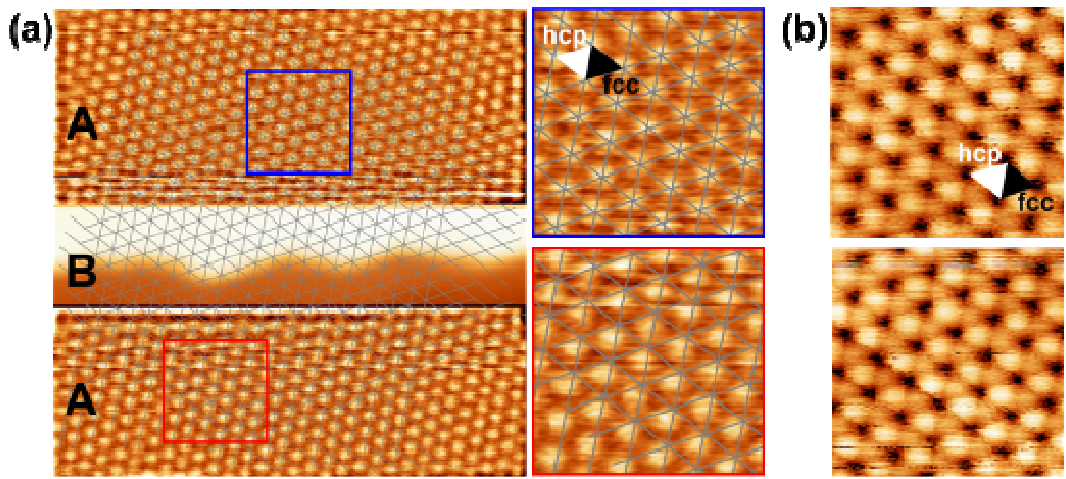

Fig. S2: (a) STM image showing an atomic contrast in both upper and lower terraces separated by a monoatomic-high step edge. The tunneling conditions were $V_s$=9.3mV

and $I_t$=100 nA in area A (on terraces) and $V_s$=9.3mV and $I_t$=1.5 nA (near the step edge). The two right panels are zoomed STM images in the blue- and red-framed area. (b) Atom manipulation images taken in the upper (top panel) and lower (bottom panel) terraces.

Figure S3 explain the procedure to obtain the STM image shown in Fig. S2(a). First, we found an area showing a monoatomic step by STM. For the atomically resolved images, we need to set the tunneling current at large values (e.g. 100 nA) but with this large current the tip may touch at the step edge during the scanning. In order to avoid the tip crash, an STM image was taken with small current (1.5 nA) around the step edge and with large current (100 nA) in terraces. After adjustment of the offset and contrast of each section in the image to observe atomic structure in both terraces and FFT filtering to eliminate the corrugation due to the moiré, we obtain the STM image shown in Fig. S2(a).

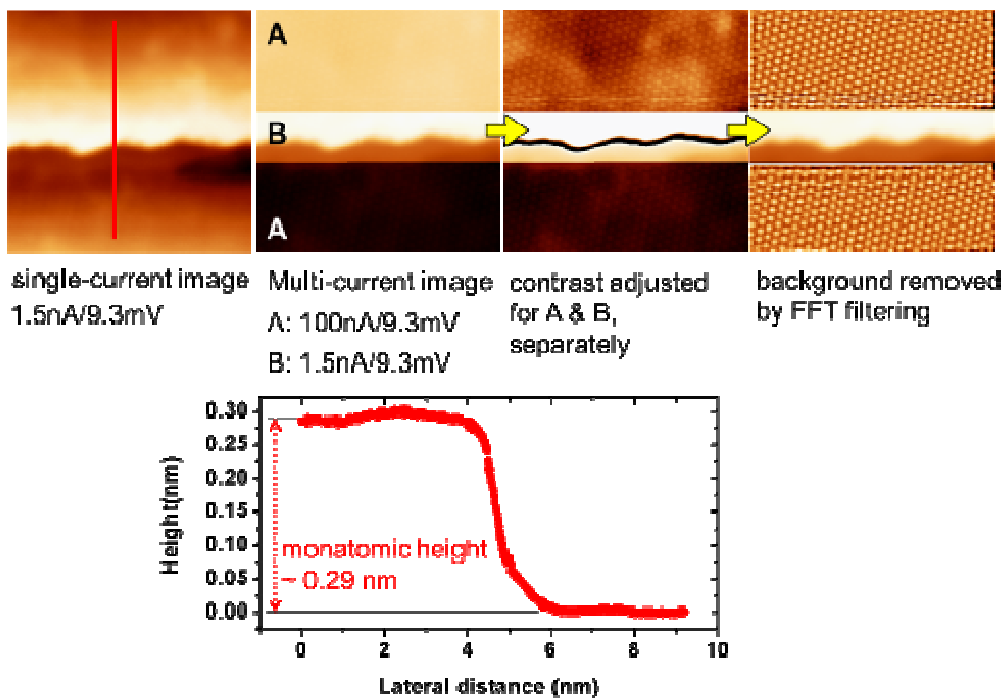

Fig. S3: Process to obtain the atomically resolved STM image across the monoatomic step shown in Fig. S2(a). (left) STM image showing a monoatomic-high step whose cross-sectional plot is shown in the bottom panel. (second left) STM image taken with 2 set currents. In area A (terraces) the current was set at 100 nA while in area B (around the step edge) it was set at 1.5 nA. (second right) The offset and contrast adjusted for atomic resolution in both upper and lower terraces. (right) FFT filtered to eliminate the corrugation due to the moiré structure.

## II. Conductance traces taken during the tip approach and retreat

Figure S4 shows the conductance traces measured at on-top and hollow sites and conductance mappings at $\Delta z$ = -32 pm and -50 pm, taken both during approaching toward the surface and during retreating from the surface. The conductance traces did not exhibit hysteresis, and the conductance mappings did not depend on the direction of the tip motion.

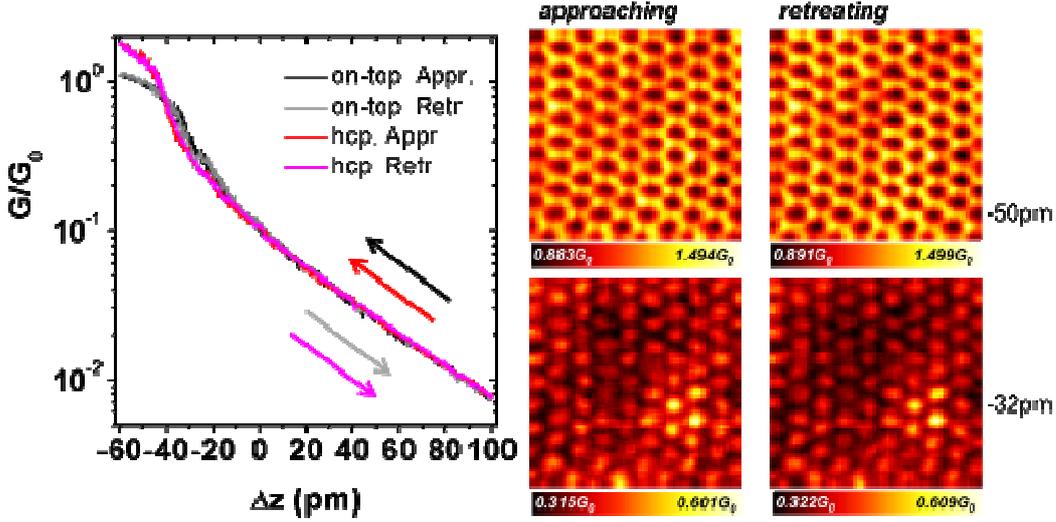

Fig. S4: (left) Conductance traces taken on-top and hcp sites during approaching and retreating of the tip to/from the surface. (right) Conductance mappings taken at -50 pm and -32 pm during the approaching and retreating. The results of the approaching are same as those shown in Figs. 1 and 2.

## III. Influence of moiré structure on the conductance

Pb islands we used in our experiments were formed on a Si(111)-$\sqrt{3}\times\sqrt{3}$ Pb reconstructed structure by Pb deposition. The azimuthal direction of the islands was rotated with respect to the Si substrate to form moiré patterns with various rotation angles and periodicities [S1-S3]. The moiré structure can be observed in topographic STM images, as shown in the STM image of Fig. S5. The moiré structure affects the conductance traces [S3] through a modification of local density of states or work function

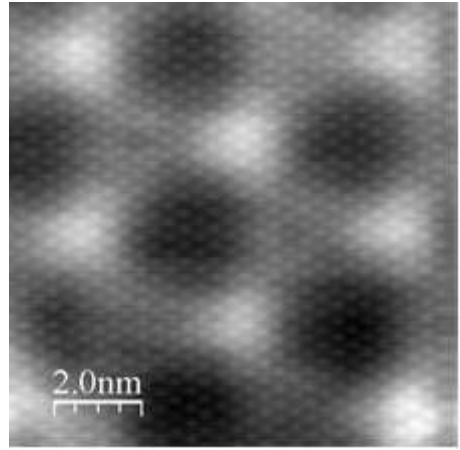

Fig. S5: an STM image (10 × 10 nm$^2$) showing moiré on top of 7ML Pb island. ($I_t$ = 30 nA and $V_S$ = 3.8 mV).

(apparent barrier height). For instance, as demonstrated in Fig. S6 the high area in the moiré structure has smaller conductance at $\Delta z$ = -60 pm, -50pm, and -32pm than the low area whereas at $\Delta z$ = 100pm the high area has larger conductance. In the analysis of the conductance traces presented in Fig. 1, we collected traces only from the high-contrasted area in the moiré in order to avoid its influence.

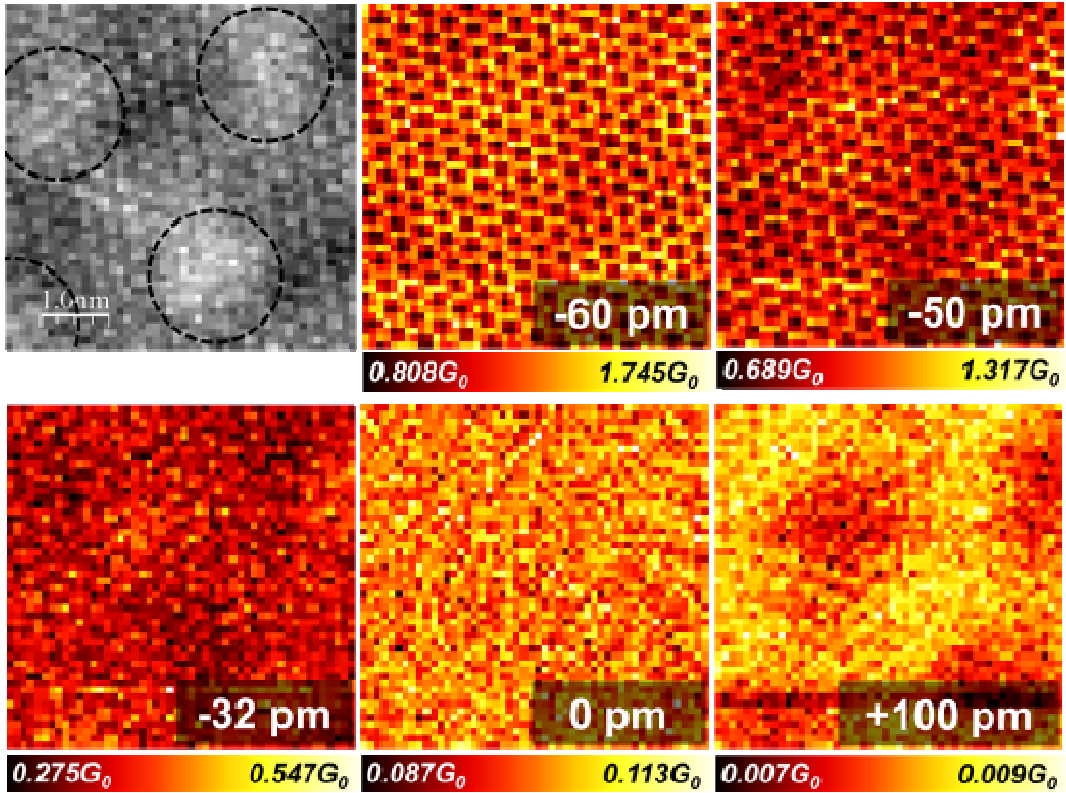

Fig. S6: (top left) STM image ($V_s$ = 3.8 mV, $I_t$ = 30 nA) showing a moiré pattern, (second left to bottom) Conductance mappings at various distances that were extracted from the simultaneously taken conductance traces over the Pb surface.

Figure S7 shows an STM image, same as Fig. 2(a), and a conductance mapping taken at $\Delta z$ = +50 pm. The area surrounded by circle A (B), which we believe corresponds to low (high) area in the moiré structure, contrasted low (high) in the $\Delta z$ = +50 pm conductance mapping. When the tip is closer to the surface ($\Delta z$ < 0), for instance, $\Delta z$ = -32 pm (Fig. 2(b)) and $\Delta z$ = -50 pm (Fig. 2(c)), the area A (B) exhibits low (high) conductance. These are consistent with the results of Fig. S6.

The bottom panel of Fig. S7 shows a plot of a ratio of the conductances taken at hcp sites and fcc sites in the low (A) and high (B) moiré areas. The plot indicates that the ratio depends not only on the displacement $\Delta z$ but also on the moiré (A or B).

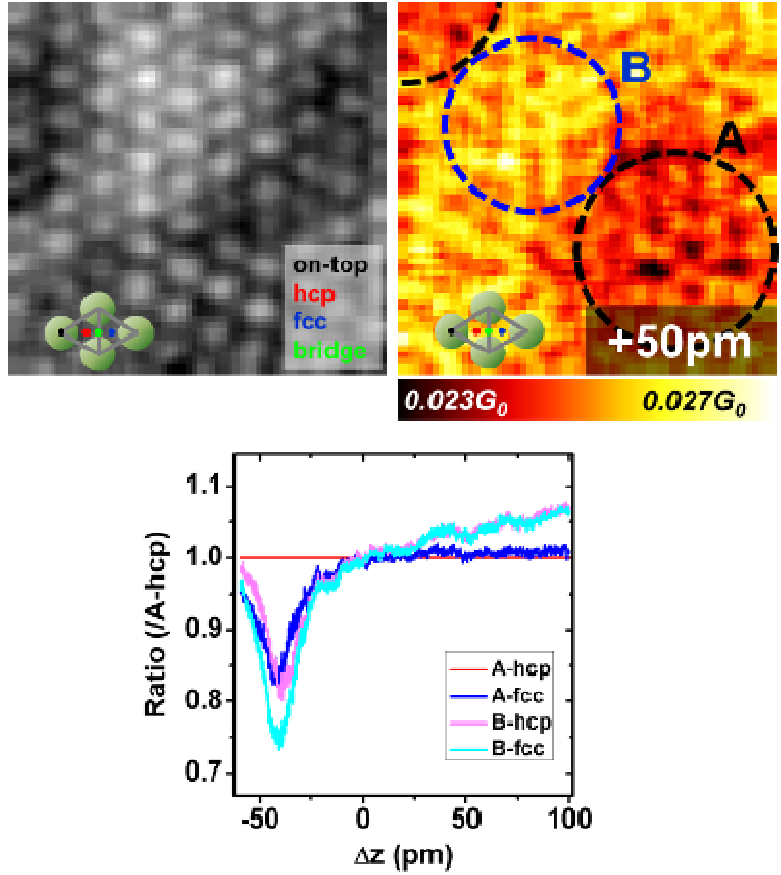

Fig. S7: (top left) STM image, same as Fig. 2(a) (top right) Conductance mapping at $\Delta z = +50$ pm. The high (low) areas in the moiré structure is circled A (B). (bottom) Ratio of the conductances taken at hcp and fcc sites in both low (A) and high (B) moiré areas

## IV. Height correction in the conductance traces

In the conductance trace measurements we forwarded the tip to the surface from the position $\Delta z=0$ set by the tunneling conditions ($V_S = 3.8$ mV, $I_t = 30$ nA for Figs. 1 and 2, and $V_S = 9.3$ mV, $I_t = 100$ nA for Fig. 3). In order to compare the conductance at the same height from the surface plane, the height difference at $\Delta z=0$ should be corrected.

In Fig. S8, we plotted the as-measured conductance traces, which are shown in Fig. 1(b), and the ones corrected by shifting the traces of the hollow (fcc and hcp) sites by -2 pm. The amount of the shift was obtained from the STM image simultaneously taken at $\Delta z=0$. Whereas overall features reported in the manuscript, such as the branching and crossover in the conductance traces, do not change significantly, the conductance values at specific $\Delta z$ are modified. The bottom panel of Fig. S8 shows a list of conductance values of $\Delta z=-50$pm and -60pm at on-top, fcc, and hcp sites, and its ratio. The panel indicates that for the quantitative analysis, as the case of Figs. 3(c) and 3(d), the height correction is required. Note that the conductance traces shown in Fig. 3(c) were corrected by shifting the trace of hollow site by -2pm.

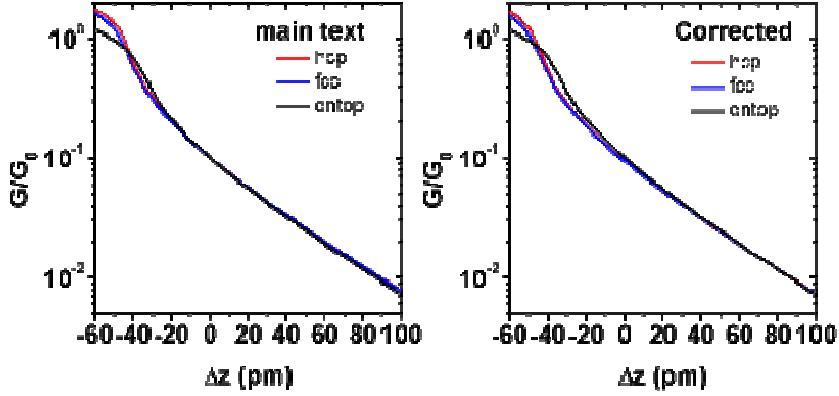

|  |  | hcp | fcc | on-top | h/t | f/t | h/t |
|---|---|---|---|---|---|---|---|
| -50 pm | raw | $1.34G_0$ | $1.21G_0$ | $0.97G_0$ | 1.38 | 1.25 | 1.11 |
|  | corrected | $1.23G_0$ | $1.09G_0$ | $0.97G_0$ | 1.27 | 1.12 | 1.13 |
| -60 pm | raw | $1.73G_0$ | $1.65G_0$ | $1.17G_0$ | 1.48 | 1.41 | 1.04 |
|  | corrected | $1.66G_0$ | $1.56G_0$ | $1.17G_0$ | 1.42 | 1.33 | 1.06 |

Fig. S8: (top left) Conductance traces taken on on-top, fcc and hcp sites (same as Fig. 1(b) of the main text). $\Delta z=0$ corresponds to the height of the tunneling condition of $V_s =$ 3.8 mV and $I_t = 30$ nA. (top right) Height-corrected conductance traces, in which the ones taken fcc and hcp sites were shifted by -2pm in order to compensate the height difference from on-top site in the set tunneling condition at $\Delta z=0$. (bottom) List of conductance values of $\Delta z=-50$pm and -60pm at on-top, fcc, and hcp sites, and its ratio

**V. Atom manipulation images**

In our experiments, we observed the larger point contact conductance at hcp site than fcc site, as shown in the conductance traces of Fig. 1(b) and the conductance mappings of Fig. 2(c). Whereas the larger contact conductance on hcp than fcc was observed for the first time experimentally in this work, the fcc-hcp contrast has also been observed in atom manipulation images, which are basically the STM images taken with a single adsorbed atom manipulated or dragged by the tip during the scanning. The fcc-hcp contrast has been demonstrated in both constant current and height image modes of the atom manipulation images.

In the manipulation images of the close-packed surface [S4-S6], the hollow sites were observed as a large triangle since the manipulated atom tends to stay there until the tip moves to a next stable hollow site. The on-top sites were observed as a small dark spot in the images. We took atom manipulation images with an adsorbed single Pb atom both in the constant current mode (Fig. S9(a), with feedback on) and in the constant height mode (Fig. S9(b), with feedback off) in a single scanning frame, and found that both modes showed the higher contrast / larger current at hcp than fcc while on-top site is darkest, as clearly demonstrated in cross-sectional plots of Fig. S9(c). The manipulated atom was transferred from the tip by mechanical contact prior to the imaging. We confirmed it was a single Pb atom by comparing normal STM images of the manipulated adatom with that reported in reference [S7].

Larger conductance on hcp site has been observed in the atom manipulation images on close-packed metallic substrates, such as Co/Cu(111) [S4] and Au/Au(111) [S5], except Ag/Ag(111), which did not exhibit contrast difference between the two hollow sites [S6]. The conditions of the manipulation images and the point contact conductance image are, however, quite different; the fcc-hcp conductance difference was observed in the atom manipulation imaging with 100 nA ($0.14G_0$) whereas with the same current STM images show normal contrast (see Fig. 3(a)). After the atom manipulation imaging the manipulated atom always remained on the surface while nothing remained after the point contact formation. Since our conductance mappings in the contact regime (Fig. 2(c)) did not exhibit triangular-shaped contrast area, unlike the case of the atom manipulation images (Fig. S9), the lateral displacement does not seem significant in the conductance trace measurements of Fig. 1(b). The situation of the manipulation images is, hence, obviously different from that of the conductance trace measurements in terms of both atomic geometry and transport.

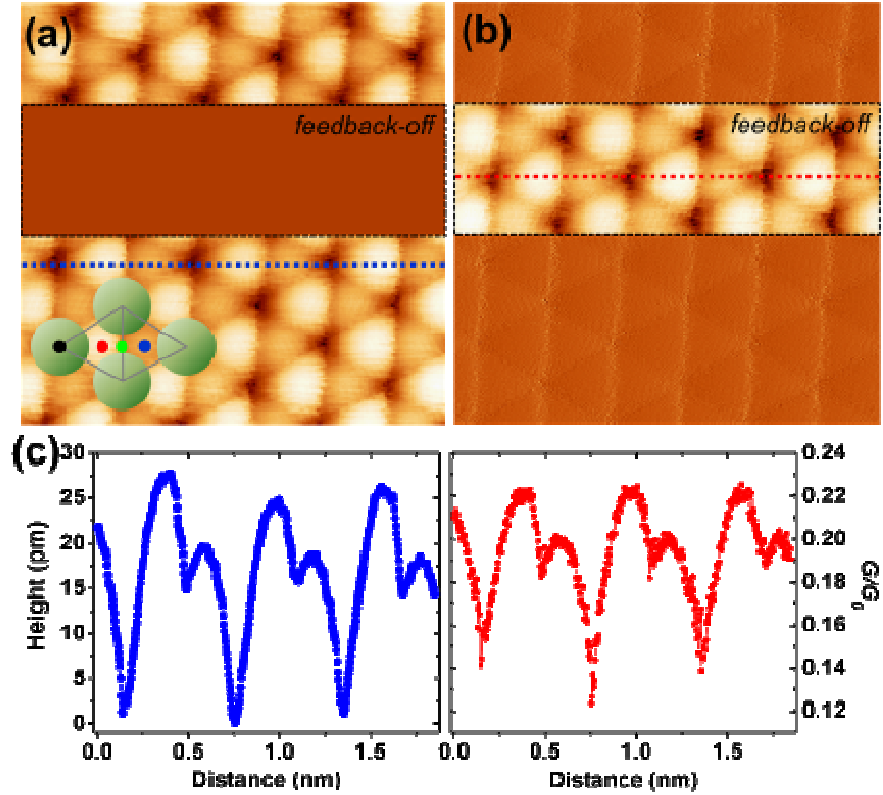

Fig. S9: (a) Constant height and (b) current mode images during atom manipulation imaging ($V_s$ = 9.3 mV, $I_t$ = 100 nA, 1.8 × 1.8 nm$^2$). In order to check the influence of the STM feedback, the feedback was turned off during the imaging. With the feedback on (off), atomically resolved image appears in height (current) images. (c) Cross-sectional plots taken along the dotted line in the height (left, blue) and current (right, red) images.

**VI. Reversible contrast in high current STM imaging**

In Fig. 3, we presented the change in atomic contrast depending on the amount of conductance. One may think that the change is due to non-reversible change in tip structure and/or electronic states. As shown in the following figure (Fig. S10), however, we confirmed that the reversed contrast in the STM image backed to normal one by setting the conductance back to low value, excluding the possibility that the observed contrast reversal is due to the non-reversible tip change. Figure 10 shows an STM image in which the set conductance was changed from $0.32G_0$ to $0.66G_0$ and then back to $0.32G_0$. Whereas the $0.66G_0$ region showed reversed contrast, the two $0.32G_0$ regions showed the same normal contrast, indicating that the observed contrast change is reversible.

The right figure shows the conductance traces taken during the approach and retreat of the tip just before the STM imaging of Fig. 3, indicating that reversible conductance traces were taken in the same tip condition as Fig. 3.

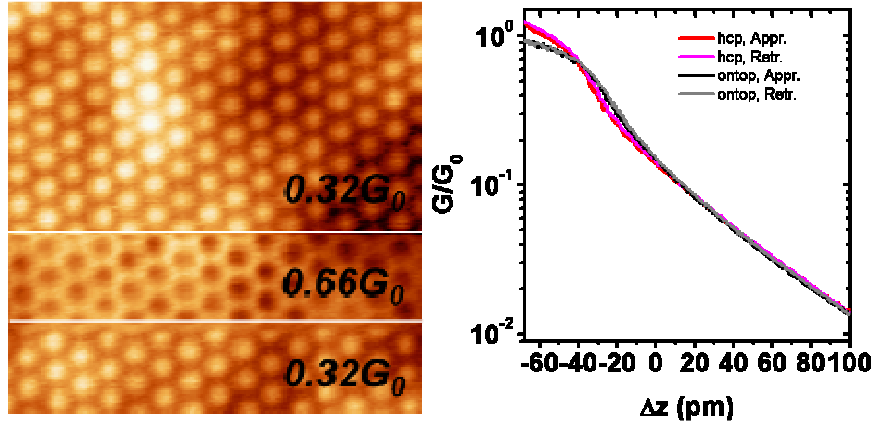

Fig. S10: (left) STM image showing normal and reversed contrasts depending on the set conductance. (right) Conductance traces taken during the tip approach and retreat just before taking STM images of Fig. 3.